\documentstyle[floats,multicol,aps,epsf,prb]{revtex}

\draft

\begin{document}

\twocolumn[\hsize\textwidth\columnwidth\hsize\csname 
@twocolumnfalse\endcsname

\title{ Instanton approach to the Langevin motion of a particle in a random potential}

\author{A. V. Lopatin$^{1}$ and V. M. Vinokur$^{2}$}
\address{$^{1}$Department of Physics, Rutgers University, Piscataway, 
New Jersey  08854\\
$^{2}$Material Science Division, Argonne National Laboratory,
Argonne, Illinois 60439}
\date{\today}
\maketitle

\begin{abstract}
  We develop an instanton approach to the non-equilibrium dynamics in 
  one-dimensional random environments. The long time behavior is controlled 
  by rare fluctuations of the disorder potential and, accordingly, by the 
  tail of the distribution function for the time a particle needs to 
  propagate along the system (the delay time). The proposed method allows 
  us to find the tail of the delay time distribution function  and delay
  time moments, providing thus an  exact description of the long-time 
  dynamics.  We analyze arbitrary environments covering different types
  of glassy dynamics: dynamics in a short-range random field, creep, and 
  Sinai's motion.  

\end{abstract}

\vskip2pc]

\bigskip 

\bigskip 

{\it Introduction--}One dimensional driven dynamics in a random 
environment attracts a great deal of current attention.  The 
motivation of the interest is two-fold: First, the propagation of a 
particle through a 1D random field has become a paradigm for a 
general  out-of-equilibrium stochastic processes in random 
systems capturing all effects of glassy dynamics including aging and 
memory effects \cite{BG,V,FV,DFPLD}.  Second, a particle  moving in a 1D 
random potential models straightforwardly a variety of physical 
systems  ranging from dislocations and charge density waves in 
solids, spin-chain dynamics and domain growth, 
to protein molecules and bacterial colonies\cite{BG,V,DFPLD,Krap,Tsim}. 
The attraction of the 1D models is thus that while 
allowing for an analytical treatment (and often even for a 
full analytical solution) they also offer an insight into generic 
basic 
properties of the wealth of glassy out-of -equilibrium systems.  
Indeed, even the simplest 1D models with the Gaussian correlated 
potential $v(x)$, with $\langle{v(x)}\rangle=0$, $\langle 
(v(x)-v(0))^2\rangle=\kappa |x|^\gamma$  exhibit a 
striking generality and diversity of glassy behaviors \cite{Scheidl,LDV}.

The approach employed in \cite{Scheidl,LDV} enables the exact 
derivation of the particle velocity, but it does not allow for a 
complete dynamic description of the system - for example, in terms of 
velocity cumulants and/or correlation functions.   Taking a kinetic 
view on the problem one can characterize the dynamic properties of 
a random system by the probability distribution for the particle 
velocity or, equivalently, by, the probability distribution for the 
particle delay time $\tau$.  The latter is defined as an average time 
that a driven particle spends to propagate through the sample, and the 
corresponding distribution function $P(\tau)$ characterizes completely 
the transport properties of the 1D system involved.  

The exact expressions for a particle velocity derived in 
\cite{Scheidl,LDV} 
indicate that the system dynamics is governed by rare 
fluctuations of the random potential, i.e. by the tail of the delay time distribution 
function.  Thus the {\it instanton} solution for the 
Langevin equation in the presence of an external force offers a most 
adequate   
description of the long time glassy dynamics in an arbitrary 1D random
environment.  In this Letter we 
apply the instanton method developed earlier in \cite{LI} to find the 
asymptotic behavior of the delay time
distribution function $P({\tau})$ at large $\tau.$
In general, the idea of the instanton method is to pick up the 
largest
contribution to the functional integral coming from an optimal 
configuration
instead of implementing the complete integration. In our case of low 
enough temperatures, the delay time is mainly 
determined by the transition over
the largest barrier in the system. Since this time is exponentially 
large,
one expects that the delay time averaged over the random potential 
(or any of its 
higher moments) is determined by the saddle point trajectory of some 
effective action controlling this exponential behavior, and, thus, 
it can be most appropriately found by 
the instanton method indeed. Further, knowing
all the moments of the delay time one can reconstruct the large $\tau$
asymptotics of the distribution function $P(\tau).$

The distribution function  $P(\tau)$ depends strongly  on the 
specific form
of the correlation function of the random potential $u(x)$. We 
restrict ourselves to the Gaussian disorder with the correlation function 
\begin{equation}
\langle v(x_1)\, v(x_2) \rangle_d=u(x_1-x_2),\;\;\;\; \langle 
v(x)\rangle=0.
\label{corr}
\end{equation}
We use also the function 
\begin{equation}
K(x)=\langle( v(0)-v(x))^2\rangle_d=2(u(0)-u(x)),  \label{kfun}
\end{equation}
which is more convenient in the case of the long-range correlated 
potential.
In the case of the short-range correlated potential, when 
$u(x)$ monotonically decreases to zero as a function
of $|x|,$ we find that  the delay time distribution is log-normal  
\begin{equation}
\ln \tilde P(Y)=-{\frac{{T^2\;Y^2}}{{4\, u(0)}}},
\end{equation}
where $Y=\ln\tau$.  The distribution function $\tilde P(Y)$ is 
defined as the
distribution function of $\ln\tau,$ being therefore related to the 
distribution
function $P(\tau)$ by 
\begin{equation}
P(\tau)={\frac{1}{\tau}}\tilde P(\ln \tau).
\end{equation}
In the case of the long-range correlated potential $K(x)=\kappa\,\, 
x^\gamma$,
with $\kappa >0$ and $0<\gamma <1$, the distribution function 
essentially
depends on the applied field $E$: 
\begin{equation}
\ln \tilde P(Y)=-{\frac{2}{\kappa }} \left[{\frac{{T\,\, 
Y}}{{2-\gamma}}} %
\right]^{2-\gamma} \left[{\frac{{E}}{\gamma}}\right]^\gamma.
\label{powerpot}
\end{equation}

And, finally, for the so-called Sinai model \cite{Sin} $K(x)=\kappa\, x$ 
 we obtain: 
\begin{equation}
P(\tau)\sim \tau^{\,-2 T E/\kappa\,\,-\,1,}
\end{equation}
which agrees with the result found earlier in Ref.\cite{FV} (see also 
\cite{BG})

{\it The model and results--} Now let us turn to the detailed 
calculations.  The dynamics of
a classical particle in a random potential is described by the 
Langevin
equation 
\begin{equation}
\Gamma^{-1} \partial_t\, x(t)=-\beta\,\partial_x v(x) +\beta E 
+\xi(t),
\label{lang}
\end{equation}
where $\beta$ is the inverse temperature, $\Gamma$ is the inverse 
relaxation time, $%
v(x)$ is the random potential, $E$ is the applied uniform field and 
$\xi(t)$ is
the Langevin thermal noise that models the thermal environment, 
\begin{equation}
\langle \xi(t_1)\, \xi(t_2) \rangle_T=2\,\Gamma^{-1}\delta(t_1-t_2).
\end{equation}

To distinguish between the average over the thermal noise and
 disorder averaging we
denote the former by the subscript $T$. The inverse relaxation time 
$\Gamma$
will be set henceforth to $1$ for convenience.  Making use of the 
standard approach (see
for example Ref.\cite{fh}) one can write the Lagrangian corresponding 
to Eq.(\ref{lang}) in a form: 
\begin{equation}
{\cal L}=-\hat x^2-i \hat x\;(\partial_t x-\beta E +\beta \partial_x 
v(x)) +{%
\frac{\beta}{2}} v^{\prime \prime}(x).  \label{lag}
\end{equation}
Further we will set $\beta=1$ measuring energies in the units of
temperature. The probability for a particle to go from the point 
$x_1$ to $x_2$
is given by the functional integral 
\begin{equation}
P(x_1,x_2)=\int D[x(t)]\;D[\hat x(t)]\; e^{\;\int dt\,{\cal L}},
\end{equation}
and the time a particle spends moving from $x_1$ to $x_2$ (the delay time) 
is, correspondingly, 
\begin{equation}
\tau(x_1,x_2) \sim P^{-1}(x_1,x_2).
\end{equation}
If the temperature is lower than a typical barrier in the system, one 
can
use the saddle point approximation to find $P$ 
\begin{equation}
P=e^A,\;\;\;\; A=\int dt\; {\cal L}_{s.p.},
\end{equation}
where ${\cal L}_{s.p.}$ is the value of the Lagrangian (\ref{lag}) at 
the saddle point trajectory.  The delay time $\tau(x_1,x_2) $ averaged over the 
disorder is then
\begin{equation}
\langle \tau(x_1,x_2)\rangle_d =\int D[v]\; e^{A[v,x]},  \label{L}
\end{equation}
with the effective action 
\begin{equation}
A[v,x]= -{\frac{1}{2}}\int dx_1 d x_2\,\, v(x_1) 
f(x_1,x_2)v(x_2)-\int dt\; 
{\cal L} ,  \label{A}
\end{equation}
where $f$ is the inverse correlation function (\ref{corr}): $\int dx\,
f(x_1,x)\, u(x,x_2)=\delta(x_1-x_2).$   Taking the variational
derivatives of the action (\ref{A}) with respect to $ \hat x,\; x, \; 
v$ we get the saddle point equations 
\begin{eqnarray}
-2\hat x+\partial_t\, x- E+\partial_x\, v(x)=0,  \label{eq1} \\
\partial_t\, \hat x-\hat x\, v^{\prime \prime}(x)=0,  \label{eq2} \\
v(x)=-\int dt\, \hat x(t) u^{\prime}(x-x(t)),  \label{eq3}
\end{eqnarray}
where the variable $\hat x$ was redefined $i\hat x \to \hat x.$ 
When deriving Eqs.(\ref{eq1},\ref{eq2},\ref{eq3}) the last term in Eq(\ref{lag})
was  neglected because it gives the ${\cal O}(1)$ contribution
to the action while the whole instanton action is much larger than one.
Following Ref.\cite{LI}, in order to find an instanton solution we 
have to choose $\hat x=\partial_t x$.
 Indeed, this allows to reduce Eqs.(\ref{eq1},\ref{eq2}) to a single
equation: 
\begin{equation}
\partial_t\, x= -E+\partial_x v(x).  \label{vel}
\end{equation}
Assuming that the instanton solution $\hat x=\partial_t\, x$ exists 
in the
interval $(\eta_1, \eta_2)$ from Eq.(\ref{eq3}) we find 
\begin{equation}
v(x)=u(x-\eta_2)-u(x-\eta_1).  \label{effpot}
\end{equation}
The interval $(\eta_1,\,\eta_2),$ where the instanton solution exists,
obviously must lie within the interval $(x_1,\,x_2)$; we will see 
later, however, that these intervals do not necessarily coincide. 
Outside the interval $(\eta_1,\, \eta_2)$ one should take the normal solution 
in the form
\begin{equation}
\hat x=0,\;\;\;\;\;\;\partial_t\, x= E-\partial_x v(x).
\end{equation}

The action (\ref{A}) corresponding to the solution 
(\ref{vel}) is
\begin{equation}
A={\frac{1}{2}}[\,v(\eta_2)-v(\eta_1)\,]- E(\eta_2-\eta_1).  
\label{act}
\end{equation}

Since within the interval $(\eta_1,\eta_2)\;\;\hat x=\partial_t x$
and outside this interval $\hat x=0,$ assuming that $\hat x$
is continuous, we conclude that on the boundaries of the interval
$(\eta_1,\eta_2)$         
$\hat x=\partial_t x=0.$ Therefore from Eqs.(\ref{vel},\ref{effpot}) we 
get the equation 
\begin{equation}
E=-u^{\prime}(\eta_2-\eta_1)  \label{sp}
\end{equation}
which defines the distance $\eta_2-\eta_1.$ Using Eq.(\ref{effpot}) 
the action (\ref{act}) can be written as 
\begin{equation}
A=u(0)-u(\eta)-E \eta, \label{finAp}
\end{equation}
where $\eta=\eta_2-\eta_1.$ Note that Eq.(\ref{sp}) corresponds to the
extremum of the action (\ref{finAp}) with respect to $\eta,$ so the 
final answer for the action can be conveniently presented as the maximal 
value of the action 
\begin{equation}
A={\frac{1}{2}} K(\eta)-E\eta  \label{finA}
\end{equation}
where $K(x)$ is given by (\ref{kfun}).

The formula (\ref{finA}) allows to find the average delay time for an arbitrary 
disorder correlation function $u(x)$  and it's concrete realization follows from the 
specific form of the correlation function $u(x)$.

(i) {\it Short range potential.} Let us begin with the simplest case 
when $u(x)$ monotonically  decreases to zero  as a function of $|x|$. The 
corresponding form
of the effective potential $v(x)$ is shown on Fig. 1. 
From Eq.(\ref{sp}) it follows that $\eta\to\infty$ when $E\to 0,$ thus
if $E$ is low enough, the action becomes 
\begin{equation}
A=u(0).   
\end{equation}

\begin{figure}[here]
\unitlength1.0cm
\begin{center}
\begin{picture}(7.5,4.2)
\epsfysize=4.2 cm
\epsfxsize=7.5 cm
\epsfbox{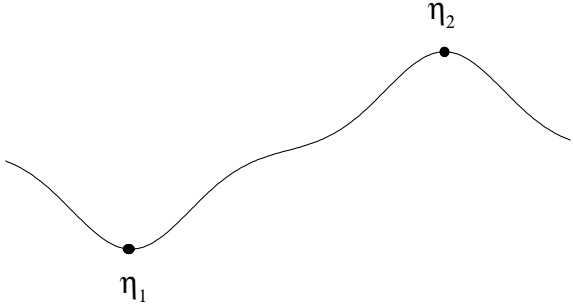}
\end{picture}
\vspace{0.6cm}
\caption{ The instanton solution for the potential $v(x)$ in the case of
short range potential. }
\end{center}
\end{figure}

(ii) {\it Potential with long-range correlations.} Let us consider the
potential of the form $K(x)=\kappa \, x^\gamma,$ with $\kappa>0$ and 
$0<\gamma <1.$ In this case the expression for the 
action (\ref{finA}) still holds; however, minimizing (\ref{finA}) with respect 
to $\eta$ we arrive at the action diverging at low fields: 
\begin{equation}
A=(1-\gamma)\, \left({\frac{ \kappa }{2}} 
\right)^{\frac{1}{{1-\gamma}}}
\left({\frac{{\gamma}}{{E }}} \right)^{\frac{{\gamma}}{{1-\gamma}}}
\end{equation}
in a full agreement with the exact solution obtained in 
Ref.\cite{LDV}

(iii) {\it `Extremely correlated' disorder, }$\gamma=1:$ This is  the 
well-known Sinai model $K(x)=\kappa\, x.$ Now
 Eq.(\ref{sp}) is either never satisfied or it is satisfied 
identically
for a special value of the applied field $E=E_0=\kappa/2$. If 
$E<E_0,$ the
instanton solution exists, but the initial and final points of the 
instanton
motion $\eta_1$ and $\eta_2$ coincide with points $x_1$ and $x_2$
respectively, and the action becomes 
\begin{equation}
A=(\kappa /2-E)\,(x_2-x_1).
\end{equation}
It means that the average time always essentially (exponentially) 
depends on
the boundary conditions. In the case $E>E_0$ the instanton solution does not exist.

{\it Distribution function of delay time.} The method described above 
can be
simply generalized for a calculation of higher moments of $\tau 
(x_1,x_2)$ 
\begin{equation}
\tau_n (x_1,x_2)=\langle \tau^n(x_1,x_2)\rangle_d.
\end{equation}
The action $A_n$ that determines the $n^{th}$ moment  
\begin{equation}
\langle \tau^n(x_1,x_2)\rangle_d = e^{A_n[v,x]}
\end{equation}
is given by 
\begin{equation}
A_n[v,x]= -{\frac{1}{2}}\int dx_1 d x_2\,\, v(x_1) f(x_1,x_2)v(x_2)-n 
\int
dt\, {\cal L}.
\end{equation}
 From this equation one can see that $A_n[u(x),x]=n\,A[\,n\, u(x),x],$
therefore using Eq.(\ref{finA}) we get the expression for the action 
$A_n$ 
\begin{equation}
A_n=n^2\, K(\eta)/2- n\, E \eta,  \label{An}
\end{equation}
that must be maximized with respect to $\eta.$ This gives 
\begin{equation}
n K^\prime(\eta)/2 -E=0.  \label{eta}
\end{equation}
Knowing all moments of $\tau$ one can find the distribution function 
$%
P(\tau).$ Indeed, in terms of the distribution function $P(\tau),$ the
moments $\tau_n$ are defined as 
\begin{equation}
\tau_n=\int P(\tau)\; \tau^n\, d \tau=\int \tilde P(Y)\, e^{nY}\,dY.  
\label{moments}
\end{equation}
Since we use the instanton method to find $\tau_n,$ with the same 
accuracy
we can use the saddle point approximation in calculation of 
the integral over $Y$ in Eq.(\ref{moments}). Comparing 
Eqs.(\ref{An},\ref{moments}) we get 
\begin{equation}
\Bigl\{\ln \tilde P(Y)+n\,Y \Bigr\}_Y =\Bigl\{n^2 K(\eta)/2- E n\, 
\eta
\Bigr\}_\eta,  \label{teta}
\end{equation}
where $\{\;\;\;\}_Y$ and $\{\;\;\;\}_\eta$ mean taking extrema with 
respect
to $Y$ and $\eta$ respectively.  Differentiating (\ref{teta}) with respect to $n$ we 
get 
\begin{equation}
n={\frac{{\ Y+E \eta }}{{K(\eta)}}}.  \label{n}
\end{equation}
Using  Eq.(\ref{teta}) we find the distribution function 
\begin{equation}
\ln \tilde P(Y)=-{\frac{1}{2}}\, {\frac{{( Y+E\eta)^2}}{{K(\eta)}}},
\label{answer}
\end{equation}
where $\eta$ is defined by the equation 
\begin{equation}
K^\prime(\eta)\,( Y+E \eta)-2\, E\, K(\eta)=0,  \label{eqeta}
\end{equation}
following from Eqs.(\ref{eta},\ref{n}). Interestingly, this equation 
also
follows form the extremum of the logarithm of the distribution 
function (\ref{answer}) with respect to $\eta.$ 

The distribution function (\ref{answer}) is our main general result. Now we shall
 analyze it's behavior for different cases introduced above:

(i) In the case of the short-range correlated potential, assuming 
that the
applied field is not too strong, from Eq.(\ref{answer}) we obtain the
log-normal distribution 
\begin{equation}
\ln \tilde P(Y)=-{\frac{{Y^2}}{{4\,u(0)}}}.
\end{equation}

(ii) In case of the potential $K(x)=\kappa\, x^\gamma,\; 0<\gamma<1$ 
from
Eq.(\ref{eqeta}) we get 
\begin{equation}
\eta={\frac{{\gamma\,\, Y}}{{E (2-\gamma)}}},
\end{equation}
and the distribution function (\ref{answer}) becomes 
\begin{equation}
\ln \tilde P(Y)=-{\frac{2}{{\kappa}}} \left[{\frac{{\ Y 
}}{{2-\gamma}}} %
\right]^{2-\gamma} \left[{\frac{{E}}{\gamma}}\right]^\gamma.
\label{powerpot}
\end{equation}

(iii) Although in the Sinai case ($\gamma=1$) the moments of $\tau$ 
are not
defined in the limit of a large system, the Eq.(\ref{powerpot}) has no
singularities when $\gamma=1$.  Therefore the distribution function 
in the Sinai
case is given by 
\begin{equation}
P(t)=t^{-2 E /\kappa\,-1},
\end{equation}
which agrees with the earlier result\cite{FV,BG}.

In conclusion, we have found the asymptotic behavior of the delay time
distribution function for a general problem of the Langevin motion of a 
particle in a
one-dimensional random potential. The application of our procedure to 
the Sinai model recovers the earlier results verifying our 
approach.  The developed instanton method 
allows to derive the distribution function for an arbitrary 1D random 
potential, and can serve as an initial step towards a quantitative 
study of the glassy dynamics of the general multidimensional systems. 
The method proposed can also be generalized for calculation of  other correlation
functions which are determined by the contribution from  the largest
barrier in the system.

{\it Acknowledgments--} AVL would like to thank Lev Ioffe for very  useful
discussions. This work was supported by the U.S. Department 
of Energy, Office of Science under contract No. W-31-109-ENG-38.

\end{document}